\documentstyle[sprocl,graphicx]{article}

\def\Journal#1#2#3#4{{#1} {\bf #2}, #3 (#4)}

\begin{document}

\hfill{ETH-TH/98-01}\\

\title{PRIMORDIAL BLACK HOLE FORMATION CAUSED BY THE QCD TRANSITION ?} 

\author{ P. WIDERIN, C. SCHMID}

\address{Institut f\"ur Theoretische Physik, 
         ETH-H\"onggerberg,\\ CH-8093 Z\"urich}


\maketitle\abstracts{
We consider the evolution of cosmological perturbations at the
QCD transition, in particular the sudden reheating from a supercooled universe
to the transition temperature. Sudden reheating happens at a specific
temperature, hence density, and singles out one specific hypersurface.  
Underdensities reach the reheating earlier than overdensities, there is a
short period of huge pressure differences which leads to a jump in the fluid
velocity. Density perturbations of scales far below the Hubble radius
$\lambda\ll R_H $ get an amplification which grows quadratically in
wavenumber. Primordial black hole formation will not be sufficiently amplified
by the QCD transition unless the initial spectrum is fine tuned.}
Could primordial black holes of one solar mass be formed because of the QCD
transition? Such black holes have been proposed as candidates for massive
compact halo objects observed by microlensing. Our analysis leads to the
conclusion that the QCD transition cannot provide sufficient amplification on
horizon scales to form black holes without fine tuning the initial spectrum. 

\section{Supercooling and Sudden Reheating}

\begin{figure}
\begin{center}
\includegraphics[width=5.9cm]{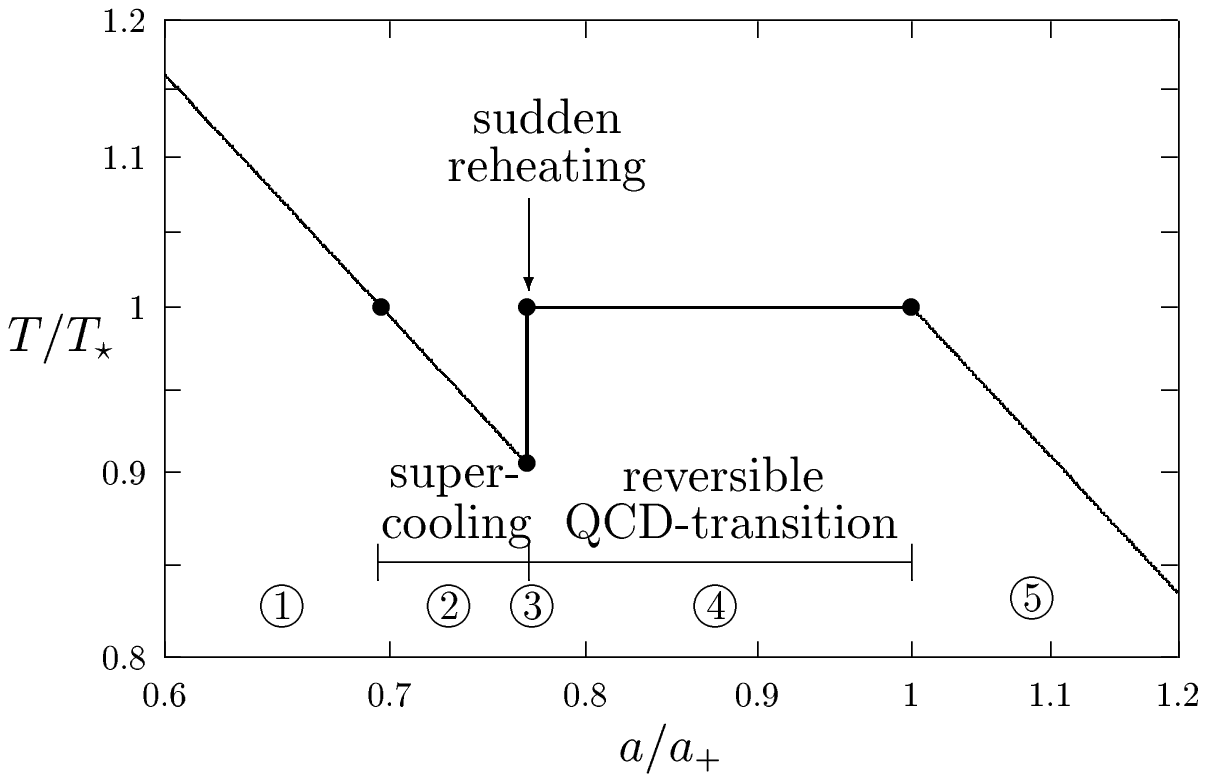}
\includegraphics[width=5.9cm]{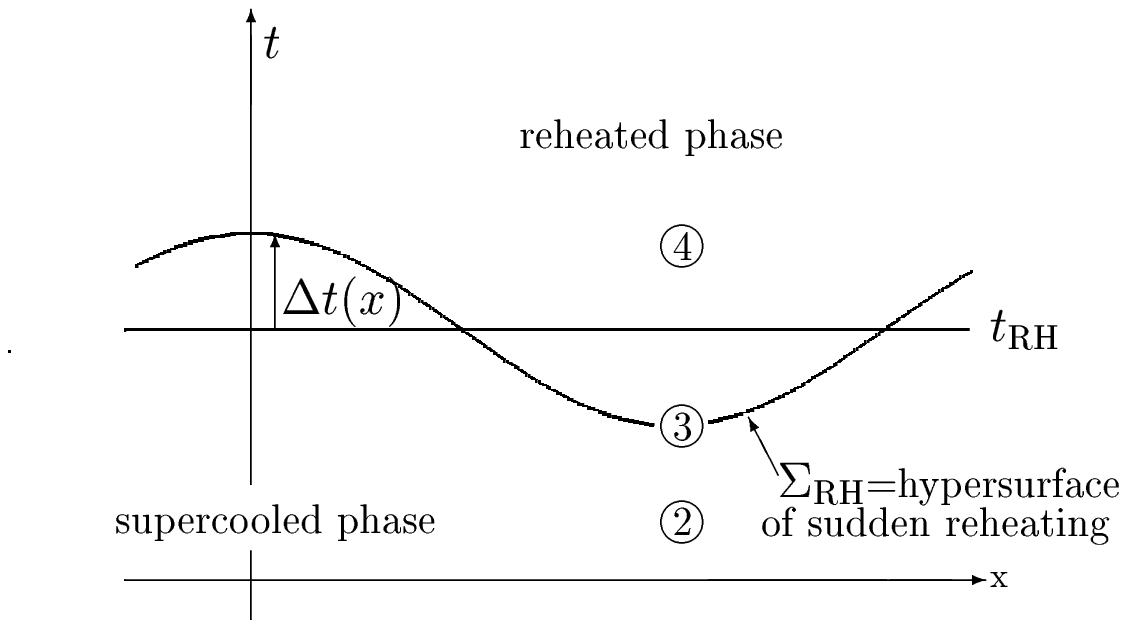}
\end{center}
\caption{\label{fig1} 
Evolution of temperature in the bag model (left) and space-time diagram
(right) at sudden reheating: The supercooled (2) and the reheated (4) epoch
are separated by the hypersurface of reheating $\Sigma_{\rm RH}$ (3). On
$\Sigma_{\rm RH}$ temperature, hence pressure, jumps uniformly. $\Delta
t(\vec{x})$ denotes the lapse of time compared to the average time, $t_{\rm
RH}$, for a certain fluid element to reach $\Sigma_{\rm RH}$. During the time
interval $O(\Delta t)$ we get large spatial variations of pressure. This leads
to huge fluid accelerations and therefore to a step in the fluid velocity. In
the general relativistic case $t_{\rm RH}$ is a hypersurface of constant
time in a certain gauge, eg. in the uniform expansion gauge.}  
\end{figure}

The QCD transition in the early universe took place at a temperature
$T_{\star} \sim 150$ MeV. Recent results of lattice QCD indicate a first order
phase transition \cite{lattice}. The quark-gluon plasma, tightly coupled to photons and leptons, cools via Hubble expansion below
$T_\star$. At the maximal supercooling \cite{Ignatius}, $\Delta T/T_\star\equiv
(T_\star-T_{\rm sc})/T_\star \approx 10^{-3}$, 
nucleation of hadronic bubbles becomes efficient and the fluid suddenly 
reheats to the transition temperature, see Fig.~\ref{fig1}. Reheating
takes only a tiny fraction of the Hubble time \cite{Ignatius}, $ \Delta t_{\rm
RH}/t_H \approx 10^{-6} - 10^{-5} $. The step in temperature implies a step in
pressure,  
\begin{equation}
  \label{deltap}
  \frac{\Delta p}{\rho+p} = \frac{\Delta T}{T_\star}.\
\end{equation}
It is crucial that this leads to large spatial variations of
pressure at the time of reheating, since underdense regions get reheated
earlier than overdense regions.  

During the reversible part of the QCD transition temperature and pressure stay
fixed and hadronic bubbles grow. There are no pressure gradients and the sound
speed is zero \cite{PRL}. After the transition is completed the hadron fluid
cools via expansion.

\section{Effects on subhorizon perturbations}

The evolution of linear cosmological perturbations is first analyzed on
subhorizon scales, where Newtonian concepts for space and time (for an
expanding radiation fluid) are applicable. Sudden reheating happens when the
maximal supercooling $\Delta T/T_\star$ is reached, i.e. on a hypersurface of
uniform temperature, hence uniform density, $\Sigma_{\rm RH}$, see
Fig.~\ref{fig1}. On $\Sigma_{\rm RH}$
pressure jumps uniformly, Eq.(\ref{deltap}). At a given Newtonian time,
underdensities have lower temperature than overdensities and therefore get
reheated earlier. The lapse of time (time delay) between the actual reheating
of a fluid element on $\Sigma_{\rm RH}$ and the average Newtonian time of
reheating, $t_{\rm RH}$, is denoted by $\Delta t(\vec{x})$ and shown in
Fig.~\ref{fig1} for one perturbation mode $k$. The time delay $\Delta
t(\vec{x})$ follows from the condition of uniform energy density on the
surface of reheating, $\rho(\vec{x},t+\Delta t(\vec{x}))=$ uniform, combined
with the continuity equation $d\rho/dt = - 3H(\rho+p)$ just before resp. just
after reheating ($t_{\rm RH}$) ,
\begin{equation}
\label{timelapse}
   \left. \Delta t(\vec{x}) = \frac{1}{3H} 
             {\delta \rho (\vec{x}) \over (\rho + p)}\right|_{t_{\rm RH}-\epsilon}
   \left.                   = \frac{1}{3H} 
             {\delta \rho (\vec{x}) \over (\rho + p)}\right|_{t_{\rm RH}+\epsilon}.
\end{equation} 
The time delay of $\Sigma_{\rm RH}$ is the same whether evaluated before or
after the hypersurface, and this condition gives the discontinuity of $\delta \rho$,
\begin{eqnarray}
  \label{jumpE}
  \left[\delta \rho \right] \equiv \delta \rho(t_{\rm RH}+\epsilon}) - \delta \rho({t_{\rm RH}-\epsilon)=
     \Delta p \left.{\delta \rho \over (\rho + p)} \right|_{t_{\rm RH}-\epsilon}.
\end{eqnarray} 
The jump in the fluid velocity is obtained by integrating the Euler equation
over the short period of huge pressure differences at $t \approx t_{\rm RH}$. 
We split $p$ into a homogeneous and an inhomogeneous term, $p(\vec{x},t) =
p(t) + \Delta p(\vec{x},t)$, where $\Delta p(\vec{x},t)$ is a sequence of step
functions with step size $\Delta p$. During the time of reheating these inhomogeneities are the dominant terms in the Euler equation 
\begin{eqnarray}
   \label{Esub}
  \partial_t \vec{S}(\vec{x},t)  &=& - \vec{\nabla} p(\vec{x},t),
\end{eqnarray} 
where $\vec{S}\equiv (\rho + p) \vec{v}$ is the momentum density. Integrating
Eq.(\ref{Esub}) over the short period of huge pressure differences gives
\begin{eqnarray}
\label{jumpS}
\left[\vec{S}(\vec{x}) \right]
            &=&\left.\Delta p \frac{1}{3 H} \vec{\nabla} {\delta \rho(\vec{x})
                \over \rho + p} \right|_{t_{\rm RH}-\epsilon}.
\end{eqnarray}
The jump conditions, Eqs.(\ref{jumpE},\ref{jumpS}), are valid for each
$\vec{k}$ mode separately. Note that the jump in $\vec{S}$ is proportional to
the wavenumber $k_{\rm phys}$. 
  
The incoming perturbations before sudden reheating are acoustic
oscillations. In the bag model the sound speed $c_s=1/\sqrt{3}$ and the
amplitudes $A_{\rm in}$ for $\delta \rho$ and $|\sqrt{3} \vec{S}|$ are equal
until sudden reheating. At $t_{\rm RH}$, $\delta \rho$ and $\vec{S}$
jump according to Eqs.(\ref{jumpE},\ref{jumpS}). After sudden reheating,
during the reversible part of the QCD transition, the sound speed $c_s \equiv
0$, i.e. the restoring force in the acoustic oscillations vanishes
\cite{PRL}. Since the QCD transition lasts less than a Hubble time, gravity is
negligible and density perturbations fall freely with constant velocity, 
$\delta \rho(t) = \delta \rho(t_{\rm RH}+\epsilon) - (t-t_{\rm
RH})\vec{\nabla} \vec{S}(t_{\rm RH}+\epsilon)$. This free fall gives an
amplification which is linear in $k_{\rm phys}$ and proportional to
$\vec{S}(t_{\rm RH}+\epsilon)$. After the
QCD transition is completed, one has acoustic oscillations again and the
total amplification factor is quadratic in wavenumber,  
\begin{equation}
\label{k1}
 {A_{\rm out}\over A_{\rm in}} 
= {k \over k_1}|{k \over k_{\rm RH}} \cos \varphi_{\rm in} - \sin
\varphi_{\rm in}| \ .
\end{equation}
$\varphi_{\rm in}$ is the phase of the incoming acoustic oscillation at $t_{\rm
RH}$. The scale $k_1 \equiv \sqrt{3} /\Delta t_{\rm trans}\approx R_H^{-1}$ 
, it depends on the duration of the reversible QCD transition $\Delta
t_{\rm trans}$, and it is due to the free fall \cite{PRL}. The scale $k_{\rm
RH}$ depends on the amount of supercooling, $k_{\rm RH} \equiv
(T_\star/\Delta T)H_{\rm RH}2\sqrt{3}$. The amplification factor is quadratic in $k$ for $(\lambda/R_H)<(\Delta T/T_{\star})$.

\section{General Relativistic Analysis}

In the general relativistic analysis of cosmological perturbations, space-time
is cut into space-like slices. For sudden reheating, there is one exceptional
slice, the hypersurface of reheating $\Sigma_{\rm RH}$, which corresponds to a
fixed-time slice in uniform density (UD) gauge \cite{Bardeen}. The extrinsic
$K_{ij}(\Sigma_{\rm UD})$ and the intrinsic curvature $^{(3)}R(\Sigma_{\rm
UD})$ remain continuous at reheating. The momentum density $\vec{S}_{\rm UD}$
stays continuous due to the momentum constraint of general relativity. Before
and after reheating we evolve in uniform expansion gauge (UE) \cite{Bardeen},
$\kappa_{\rm UE} \equiv \delta tr[K_{ij}(\Sigma_{\rm UE})] \equiv 0$. We now
derive the jump conditions in UE gauge. Just before reheating we gauge
transform to UD gauge, join and transform back to UE gauge just after reheating. The momentum density is expressed in terms of a scalar potential, $\vec{S}\equiv
\vec{\nabla}\psi$. The required gauge transformations are
\cite{Bardeen} \noindent
\begin{eqnarray}
\label{KGauge}
\tilde{\kappa} - \kappa&=& - (3\dot{H}  - k_{\rm phys}^2)  \Delta t \\
\label{PGauge}
\tilde{\psi} - \psi &=& - (\rho_0 + p_0) \Delta t .
\end{eqnarray}
$\Delta t$ is the lapse of time from the old to the new hypersurface
(gauge). The lapse of time from $\Sigma_{\rm UE}$ to $\Sigma_{\rm UD}$ is
derived in the same way and with the same result as Eq.(\ref{timelapse}). $\kappa_{\rm UD}$ and
$\psi_{\rm UD}$ are expressed in terms of $\delta \rho_{\rm UE}$ and
$\psi_{\rm UE}$ using Eqs.(\ref{KGauge},\ref{PGauge}). Since
$\left[\kappa_{\rm UD} \right]=\left[\psi_{\rm UD} \right]\equiv 0$ at the
time of sudden reheating we obtain the general relativistic jump conditions at
sudden reheating in UE gauge:  
\noindent
\begin{eqnarray}
\label{jump}
\left[\delta \rho_{\rm UE} \right]=3 H \left[\psi_{\rm UE} \right] =
      \Delta p  
      {k_{\rm phys}^2\over k_{\rm phys}^2 - 3 \dot{H}}
      \left.{\delta \rho_{\rm UE} \over \rho + p}\right|_{t_{\rm RH}-\epsilon}
\end{eqnarray}
For superhorizon scales, $k_{\rm phys} \ll H$, the right hand side becomes
negligible and we get no effect. In the subhorizon limit $k_{\rm phys} \gg H$
we recover Eq.(\ref{jumpE},\ref{jumpS}).
 
We conclude that very large amplifications of perturbations are produced for
scales $\lambda$ far below the horizon. But primordial black holes production
needs nonlinear perturbations at the horizon scale. This cannot be produced by
amplifications due to the QCD transition even with supercooling.

\section*{Acknowledgments}
We would like to thank Dominik Schwarz for useful discussions. P. W. thanks
the Swiss National Science Foundation for financial support.

\end{document}